\documentstyle[amsfonts,12pt]{article}

\newcommand{\h}{{\hbar}}

\newcommand{\de}{\delta}

\newcommand{\la}{\lambda}

\newcommand{\ve}{\varepsilon}

\begin{document}

\vspace{0.3in}
\begin{flushright}
 ITEP-TH-30/02\\
hep-th/0206039\\
\end{flushright}
\vspace{10mm}

\begin{center}
{\bf ON THE FEDOSOV DEFORMATION QUANTIZATION BEYOND THE REGULAR POISSON
MANIFOLDS}\\[1cm]
V.A. Dolgushev$^{a,b}$, A.P. Isaev$^{a,c}$, S.L. Lyakhovich$^{d,e}$
and A.A. Sharapov$^e$
\end{center}

\begin{center}
{\it $^a$ Bogoliubov Laboratory of Theoretical Physics, JINR, Dubna,} \\
{\it 141 980, Moscow Region, Russia} \\
{\it $^b$ Institute for Theoretical and Experimental Physics, } \\
{\it 117259, Moscow, Russia} \\
{\it $^c$ Max-Plank-Institute f\"{u}r Mathematik} \\
{\it Vivatsgasse 7, D-53111 Bonn, Germany } \\
{\it $^d$ Department of Theoretical Physics, Chalmers University of Technology,} \\
{\it S-412 96 G\"oteborg, Sweden} \\
{\it $^e$ Department of Theoretical Physics, Tomsk State University, }\\
{\it 634050, Lenin av. 36, Tomsk, Russia}
\end{center}

\begin{abstract}
A simple iterative procedure is suggested for the deformation
quantization of (irregular) Poisson brackets associated to the classical
Yang-Baxter equation. The construction is shown to admit a pure algebraic
reformulation giving the Universal Deformation Formula (UDF)
for any triangular Lie bialgebra.
A simple proof of classification theorem for
inequivalent UDF's is given.
As an example the explicit quantization formula is
presented for the quasi-homogeneous Poisson brackets on two-plane.

\end{abstract}
%\thanks{}

% \date{September  2000}
% \subjclass{}

% \dedicatory{}
% \thanks{}

\section{Introduction}

The Deformation Quantization, as it was originally formulated in
\cite{Berezin,BFFLS}, has undergone an explosive development
during the last two decades \cite{Stern} with a number of
important achievements including Kontsevich's $\ast$-product
construction for a general Poisson manifold \cite{Kontsevich}.
Fedosov \cite{Fedosov,Fedosovbook} gave a simple and manifestly
covariant quantization method working well for any symplectic or
regular Poisson manifold and yet giving a simple classification
for inequivalent quantizations.  Meanwhile, this method can not
be directly applied to the irregular Poisson brackets because the
main ingredient is lacking for the Fedosov construction:  no
affine connection can exist respecting an irregular Poisson
structure.

In this paper we suggest a simple method extending the Fedosov
approach to a broad class of manifolds with irregular Poisson
brackets, that gives an explicitly covariant and geometrically
transparent algorithm for constructing the $\ast$-products in the
irregular case. To make the method working, these Poisson
manifolds are assumed to be equipped with certain additional
algebraic structures. We also expect that the method will allow for
generalizations essentially relaxing conditions imposed on the
class of Poisson manifolds. Although the brackets of considered
class, being more general than the regular ones, are of interest
in their own rights, our approach admits also a purely algebraic
reformulation, which allows one to simply derive the Universal
Deformation Formula for any triangular Lie bialgebra (see Sec. 5).

To explain the class of the irregular brackets we are going to
study, let us give a simple example of this type considered in
\cite{R93,R98}.

Consider a collection of pair-wise commuting vector fields $X_i$,
$ i=1,...,n$ defined on a smooth real manifold $M$. These fields
may be viewed as derivatives of the commutative algebra of smooth
functions $ C^{\infty}(M)$. A constant skew-symmetric matrix $r$
assigns $M$ with a Poisson bracket of the form
\begin{equation}  \label{RB}
\{f,g\}=r^{ij}(X_if)(X_jg)\,,\quad\quad \forall\, f,g\in
C^{\infty}(M)\,.
\end{equation}
In general, this bracket is irregular since the rank of the distribution $
X_i $ can vary from point to point. Nevertheless, it can be
easily quantized by the Weyl-Moyal like formula
\begin{equation}
f*g=f\cdot g+\sum_{k=1}^{\infty} \left(-\frac{{i\hbar}}{2}\right)^k
\frac{1}{k!}
r^{i_1 j_1}\ldots r^{i_k j_k} (X_{i_1} \ldots X_{i_k} f)(X_{j_1}
\ldots X_{j_k} g)\,, \label{gen-n}
\end{equation}
${\hbar}$ being a formal deformation parameter. The associativity of the $\ast$
-product follows trivially from the commutativity of vector fields $X_i$.
Note that (\ref{gen-n}) may be regarded as the {\it universal deformation
formula}, in the sense that it works for the action of {\it any} commutative
Lie algebra.

A natural generalization of the above construction, we are going to study
in this paper, consists in allowing the vector fields $X_i$ to form a
noncommutative Lie algebra, say
\begin{equation}
[X_i,X_j]=f_{ij}^kX_k\,,  \label{lalg}
\end{equation}
$f_{ij}^k$ being structure constants. The Jacobi identity for the Poisson
bracket (\ref{RB}) implies that
\begin{equation}  \label{JI}
\{f,\{g,h\}\}+cycle(f,g,h)=\Lambda ^{ijk}(X_if)(X_jg)(X_kh)=0\,,
\end{equation}
where
\begin{equation}
\Lambda ^{ijk}=f_{mn}^ir^{mj}r^{nk}+cycle(i,j,k)\,.
\end{equation}
For the commutative algebra we have $\Lambda \equiv 0$, and the
Jacobi identity is automatically satisfied.  In general case,
equation $\Lambda =0$ is rather nontrivial and known in
mathematical literature as the {\it  classical Yang-Baxter
equation} (CYBE).  For reasons, which we will not explain here, a
skew-symmetric matrix $r$ satisfying this equation is called the
{\it classical triangular $r$-matrix} \cite{Dr83}.  If all the
vector fields $X_i$ are linearly independent (at least at one
point on $M$), the CYBE provides both necessary and
sufficient conditions for the Jacobi identity to hold.  In the
opposite case, the equation $\Lambda = 0$ gives only sufficient
condition. The important observation concerning $r$-bracket
(\ref{RB}) is that one may always assume the $r$-matrix to be
non-degenerate without restricting generality. Indeed, as for any
skew-symmetric matrix $r$, one may always find such a basis of
the generators $X_i=(X_A,X_\alpha )$ in which $r^{i\alpha }=0$
and ${\rm \det }(r^{AB})\neq 0$. Thus, only the vector fields
$X_A$ actually contribute to the Poisson bracket (\ref{RB}). From
the CYBE it then follows that
\[
\Lambda ^{\alpha AB}=f_{MN}^\alpha r^{MA}r^{NB}=0\quad \Rightarrow \quad
f_{MN}^\alpha =0
\]
and hence, the vector fields $X_A$ form a Lie subalgebra, for
which $ r^{AB}$ is a non-degenerate triangular $r$-matrix. For
non-degenerate $r$ -matrices the CYBE reduces to the ordinary
cocycle condition
\begin{equation}  \label{cc}
f^n_{ij}r_{nk}+cycle(i,j,k)=0\,,
\end{equation}
where $ r_{ik}r^{kj}=\de^j_i. $ The solutions to the equation
(\ref{cc}), forming a linear space of 2-cocycles, are known to be
in one-to-one correspondence with the central extensions of the
Lie algebra. In other words, equation (\ref{cc}) arises as a part
of the Jacobi identity for the Lie algebra
\begin{equation}  \label{fla}
[y_i,y_j]=f_{ij}^ky_k+r_{ij}c\,,\quad\quad [c,y_i]=0\,
\end{equation}
The Lie algebra admitting a central extension defined by a
non-degenerate cocycle $r_{ij}$ is called {\it quasi-Frobenius},
and {\it Frobenius} if at least one of such cocycles is trivial
(representable in the form $r_{ij}=f^k_{ij}\xi_k$ for some vector
$\xi_k$)\cite{El}. An extended list of examples of
(quasi-)Frobenius Lie algebras  (from here on QFL algebras, for
short) can be found in recent papers \cite{L1}, \cite{L2},
\cite{L3}.

In this paper we give a simple quantization formula for the
Poisson bracket (\ref{RB}) associated to the action of an
arbitrary QFL algebra (\ref{lalg}). Our method, being
conceptually similar to that of Fedosov, does not require
regularity of the bracket (\ref{RB}, \ref{lalg}). The main
distinction is a possibility  to work with an auxiliary quantum
bundle associated to the enveloping of the underlying Lie
algebra, instead of the usual quantum Weyl algebra  exploited in
the original Fedosov construction and its various adaptations and
reinterpretations \cite{BW}, \cite{GL}, \cite{BGL}, \cite{DLS},
\cite{KS}. Since the output $\ast$-product involves only initial
algebraic data (much like its commutative version (\ref{gen-n}))
one may regard it as the Universal Deformation Formula for the
respective algebra.

Finally, we should remark that the quantization problem for the
$r$-brackets  is deeply rooted to the theory of quantum groups
\cite{Takh}, dating back to the seminal Drinfeld's paper
\cite{Dr83}, where the existence of quantization was proved for
an arbitrary triangular $r$-matrix. Actually, the computations
required for obtaining explicit expressions by the Drinfeld
method are appeared to be impracticable except for the Abelian
case. That is why the most of known examples of the deformation
has been constructed by means of {\it a twisting} transformation
technique rather than by the quantization \cite{L1},
\cite{L2}, \cite{L3}, \cite{UDF}. In the recent paper \cite{XU}, the Fedosov
method has been used to quantize a non-degenerate triangular {\it
dynamical} $ r$-matrices. In the formal algebraic settings the
application of the Fedosov deformation quantization to a certain
class of irregular Poisson brackets, including $r$-brackets
(\ref{RB}), (\ref{lalg}) has been also discussed in \cite{V}. In
principle, the method of the work \cite{XU}, as well as it's
predecessor \cite{Dr83}, gives rise to the quantization of the
$r$-bracket (\ref{RB}), and even more general Poisson structures.
At the technical level, however, this method appears to be unduly
cumbersome for the application to such a simple class of Poisson
brackets as we are going to consider. In this respect, our
approach, being elementary in essence, will hopefully be found
useful both from theoretical and practical viewpoints.

The structure of the paper is as follows: The deformation quantization
of general $r$-brackets is exposed in Section 2. Section 3 is devoted
to the classification of such quantizations and some possible modifications
of our method.
In Section 4, we consider the case of quasi-homogeneous Poisson
brackets on two-plane which is shown to provide an interesting class of
$r$-brackets associated with the two-dimensional
(and therefore quasi-Frobenius) Lie algebras.
In Section 5, we show how the construction of the
Universal Deformation Formula for a triangular $r$-matrix results
in a quantization of the respective Lie bialgebra.
In concluding Section, we summarize the results and sketch proposals
for further generalizations of our construction within the
framework of BRST quantization.

\section{Construction of the $\ast$-product}

Let us start with the action $\rho : L\rightarrow {\rm Vect}(M)$
of an abstract QFL algebra $L$ on a smooth real manifold $M$.
Denote by $L_c$ the central extension of $L$ associated with a
non-degenerate 2-cocycle $r$. We assume that upon choosing a basis
the structure of the algebra $L_c$ is described by the Lie
brackets (\ref{fla}), so that $L=L_c/ {\Bbb R}c$ and the vector
fields $\rho (y_i)=X_i$ commute according to (\ref {lalg}).

Introduce the associative algebra ${\cal A}=C^\infty (M)\otimes
U(L_c)$, being a tensor product of the commutative algebra of
smooth functions on $M$ and the universal enveloping of the Lie
algebra $L_c$ over ${\Bbb C}$ (or more precisely, its formal
completion by infinite series). The elements of ${\cal A}$ may be
viewed as the sections of the trivial vector bundle $p
:U(L_c)\times M\rightarrow M$, with the fiber-wise associative
algebra structure induced by the associative multiplication in
$U(L_c)$. Choosing the basis of monomials being symmetric in the
Lie algebra generators, we identify the elements of $U(L_c)$ with
their Weyl symbols and set $c=1\in{\Bbb C}$. Then the generic
element $a\in {\cal A}$ reads
$$
a(y)=\sum_{k=0}^\infty a^{i_1\cdots i_k}y_{i_1}\cdots y_{i_k}\,,
$$
where $y_i$ are formal {\it commuting} variables and the
coefficients $ a^{i_1\cdots i_k}\in C^\infty (M)$ are symmetric
in the indices $ i_1,\ldots ,i_k$. The product of two symbols
$a,b\in {\cal A}$, which we will denote by $\circ $, is given by
\footnote{See \cite{KM} for a detailed survey of the operator
calculus, in particular, for the Weyl calculus on Lie algebras.}
\begin{equation}\label{circ}
(a\circ b)(y)= a(\hat{L})b(y)\,, \quad \quad \hat{L}_i=
 \left(y_j+\frac12r_{jn}\frac{\partial}{\partial y_n}\right){\cal
R}^j_i\left(\frac{\partial}{\partial y_k}\right) \,,
\end{equation}
where
\begin{equation}\label{}
{\cal R}(x)= \sum_{m=0}^{\infty}\frac{b_m}{m!}\Lambda ^m(x)
=\left(\frac{e^{\Lambda (x)}- 1}{\Lambda
(x)}\right)^{-1}\,,\qquad \Lambda^i_j(x)=x^kf_{kj}^i\,,
\end{equation}
$b_m$ being Bernoulli numbers. The formula involves well-known
group-theoretical constructions: matrices $\Lambda(x)$ realize
the adjoint representation of the Lie algebra $L$; ${\cal R}(x)$
are the matrices of the right shifts on the group $G={\rm
Exp}(L)$, written in the first kind coordinates $x^i$; finally,
operators $\hat{L}_i$ define the left regular representation of
the algebra $L_c$ corresponding to the Weyl ordering \cite{KM},
\begin{equation}
  [\hat{L_i},\hat{L_j}]=f_{ij}^k\hat{L}_k+r_{ij}\,.
\end{equation}
In what follows the explicit form (\ref{circ}) of the
$\circ$-product will be inessential for our considerations.

The space of smooth functions $C^{\infty}(M)$ is embedded into
${\cal A}$ as a central subalgebra.  Denote by $\pi:  {\cal
A}\rightarrow C^{\infty}(M)$ the canonical projection $(\pi
a)(y)=a(0)$.  To assign $C^{\infty}(M)$ with a $\ast$-product
compatible with the Poisson brackets (\ref{RB}) we construct
another, less trivial embedding $\sigma: C^{\infty}(M)\rightarrow
{\cal A}$, allowing one to induce the desired $\ast$-product via
pull-back of $\circ$-product.  To this end, introduce the
following Lie algebra of external differentiations:
\begin{equation}
\label{der}
\begin{array}{ll}
D_ia=X_ia+[y_i,a]\,,&\\[3mm]
D_i(a\circ b)=(D_ia)\circ b+a\circ (D_i b)\,,  &\forall\, a,b \in
{\cal {A }} \,,\\[3mm]
[D_i,D_j]=f_{ij}^kD_{k}\,.&
\end{array}
\end{equation}
Hereafter we set $[a,b]=a\circ b - b\circ a$.  Denote by ${\cal
A}_D$ the subalgebra of $D$-constant elements
\begin{equation}  \label{AD}
{\cal A}_D=\{a\in {\cal A}|\,\, D_ia=0\,\}\,.
\end{equation}
The following theorem establishes isomorphism between the linear spaces $
C^\infty(M)$ and ${\cal A}_D$.

\vspace{0.5cm}

\noindent {\bf Theorem 1.} {\it Any $D$-constant element $a\in
{\cal A}_D$ is uniquely determined by its projection $\pi a$ onto
subspace $C^{\infty}(M)$ of $y$-independent elements and vice
versa.}

\vspace{0.5cm}

\noindent {\it Proof.} Let
\begin{equation}  \label{exp}
a(y)=\sum_{n=0}^\infty a_n\,,\quad\quad a_n=a^{i_1\dots
i_n}y_{i_1}\dots y_{i_n}\,,
\end{equation}
be an arbitrary element from ${\cal A}$ with a given $a_0=a(0)\in
C^{\infty}(M)$. Expanding the condition $D_ia=0$ via the basis of
homogeneous monomials we get a chain of equations
\begin{equation}
\partial_i a_{n+1}=V_i a_{n}\,,  \label{oni}
\end{equation}
where we have denoted
\begin{equation}
\partial_i=r_{ji}\frac{\partial}{\partial y_j}\,,\quad\quad
V_i=X_i+y_kf^k_{ij}\frac{\partial}{\partial y_j}\,.
\end{equation}
Besides, the system (\ref{oni}) should be supplemented by a set of
compatibility conditions resulted from the commutativity of the
partial derivatives $\partial_i$,
\begin{equation}  \label{ccon}
\partial_i(V_j a_{n})-\partial_j(V_i a_{n})=0\,.
\end{equation}
To construct a solution to equations (\ref{oni}), (\ref{ccon}) we
use the induction over the degree $n$ in the expansion
(\ref{exp}). For $n=0$ the equations are obviously consistent and
we get
\begin{equation}
a_{1}=y_ir^{ji}X_ja_{0}\,,\quad a_{0}\in C^{\infty}(M)\,.
\end{equation}
Suppose that we have found all $a_k$ with $k=1,2,...,n$. Then
equations (\ref{oni}) for $a_{n+1}$ are compatible provided the
conditions ($\ref{ccon}$) are satisfied. Using identities
\begin{equation}  \label{id}
[\partial_i, V_j] -[\partial_j, V_i] =f_{ij}^k\partial_k\,,\quad\quad
[V_i,V_j]=f_{ij}^kV_k\,,
\end{equation}
we can rewrite (\ref{ccon}) as
\begin{equation}
\partial_i(V_j a_{n})-\partial_j(V_i a_{n}) =f^k_{ij}\partial_k a_{n}-
(V_i\partial_j-V_j\partial_i)a_{n}\,,
\end{equation}
and since by induction hypothesis $\partial_ia_{n}=V_ia_{n-1}$ we
finally get
\begin{equation}
f_{ij}^k\partial_ka_{n}-[V_i,V_j]a_{n-1}=f_{ij}^k( \partial_ka_{n}-V_k
a_{n-1})=0\,.
\end{equation}
Upon the compatibility conditions have satisfied the
straightforward integration of (\ref{oni}) leads to the explicit
recurrent formula
\begin{equation}  \label{iter}
a_{n+1}(y)=\int_0^1y_jr^{ij}\left(X_ia_{n}(ty)+y_kf^k_{im} \frac{
\partial a_{n}(ty)}{\partial y_m}\right)dt
\end{equation}
for the unique $D$-constant element $a\in {\cal A}_D$ with a
prescribed $ a_0=a(0)\in C^{\infty}(M)$. q.e.d.

Up to the second order in $y$ the lift of an arbitrary function $a\in
C^{\infty}(M)$ to a $D$-constant element of ${\cal A}$, denoted by $
\sigma(a) $, reads
\begin{equation}  \label{lift}
\sigma(a)=a+y^iX_ia+\frac12(y^iy^jX_iX_ja+y_jy^if_{in}^jX^na)+\cdots
\end{equation}
$$
y^i=r^{ji}y_j\,,\,\,\,\,\,\,X^i=r^{ji}X_j\,.
$$

As the next step we introduce the deformation parameter $\hbar$ and extend
the space of smooth functions, i.e. the space of classical observables, to
that of quantum observables $C^{\infty}(M)[[\hbar]]$. The latter consists of
elements
\[
a(\hbar)=\sum_{k=0}^{\infty}\hbar^k a_k\,, \quad \quad a_k\in
C^{\infty}(M)\,,
\]
being formal power series in $\hbar$ with coefficients in smooth functions.
The above definitions of $\circ$-product and derivatives (\ref{der}) are
extended by linearity to the algebra ${\cal A}[[\hbar
]]=C^{\infty}(M)[[\hbar ]]\otimes U(L_c)$; in so doing, we rescale the
structure constants of the algebra $L_c$ by multiplying on $i\hbar$,
\begin{equation}  \label{rsc}
f^k_{ij}\rightarrow i\hbar f^k_{ij}\,,\quad\quad r_{ij}\rightarrow i\hbar
r_{ij}\,,
\end{equation}
and redefine the action of the derivatives
\begin{equation}  \label{rd}
D_i\rightarrow D_i=X_i+\frac{1}{i\hbar}[\,y_i,\,\cdot\,]\,,
\end{equation}
so that the lifting map (\ref{iter}) remains intact. The relation
of the above theorem to the quantization of a general $r$-bracket
(\ref{RB}) is established by the following

\vspace{5mm}

\noindent {\bf Theorem 2}. {\it The pull-back of $\circ$-product
on ${\cal A}[[\hbar]]$ via ${\Bbb C}[[\hbar]]$-linear isomorphism
$\sigma: C^{\infty}(M)[[\hbar]]\rightarrow {\cal A}[[\hbar]] $
induces an associative $\ast$-product on
$C^{\infty}(M)[[\hbar]]$},
\begin{equation}
\label{star}
a\ast b=\pi(\sigma (a)\circ \sigma(b))=\sum_{k=o}^{\infty}\hbar^k
D_k(a,b)\,
\end{equation}
{\it satisfying the properties}

\vspace{2mm}

i) {\it Locality : $D_{k}(a,b)$ are bi-differential operators of
order $k$;}\hfill (25.$a$)

\vspace{2mm}

{\it ii) Correspondence principle:} $a\ast b=a\cdot
b-\displaystyle \frac{i\hbar}{2}\{a,b\}\,\,\,({\rm
mod}\,\hbar^2)$;\hfill (25.$b$)

\vspace{2mm}

{\it iii) Normalization condition:} $1*a=a*1=a$.\hfill (25.$c$)

\vspace{5mm}

\noindent {\it Proof}. The associativity of $\ast$-product
directly follows from associativity of $\circ$-product and the
identity $\sigma \pi |_{{\cal A}_D}={\rm id}$. To prove the
locality condition it is convenient to assign the algebra ${\cal
A}[[\hbar ]]$ with a filtration \[ {\cal A}[[\hbar ]]={\cal
A}_0\supset {\cal A}_1\supset {\cal A}_2\supset \cdots
\] associated to the sequence of subspaces ${\cal A}_k$ consisting of
elements \[ a=\sum_{2m+n\geq k}\hbar ^ma_m^{i_1\dots
i_n}y_{i_1}\dots y_{i_n}\in {\cal A}_k\,.  \] This filtration
defines the topology and convergence in the space of formal power
series in $\hbar $ and $y$'s. From the recurrent formula
(\ref{iter}) we see that the element $a_n$ belongs to ${\cal
A}_n$ and involves {\it finite} linear combinations of $a_0$ and
its derivatives along $X_i$ up to order $n$. In view of the
obvious inclusions (keep in mind redefinitions (\ref{rsc}))
\[
{\cal
A}_m\circ {\cal A}_n\subset {\cal A}_{n+m}\,,\quad \quad \pi ({\cal A}
_n)\subset {\cal A}_n
\]
we can write
\[
a*b=\sum_{n+m\leq 2k}\pi (a_n\circ b_m)\quad ({\rm mod}\,\,\hbar ^{k+1})
\]
and hence, each power of $\hbar $ in the product expansion is
given by a finite order bi-differential operator. More accurate
consideration of the above formula shows that
$$
D_k(a,b)=\frac {(-i)^k}{2^kk!}r^{i_1j_1}\ldots
r^{i_kj_k}(X_{i_1}\ldots X_{i_k}a)(X_{j_1}\ldots X_{j_k}b)+\cdots\,,
$$
where dots stand for the bi-differential operators of order less
than $k$. For the commutative Lie algebra $L$ the rest terms
vanish and we come to the Weyl-Moyal $\ast$-product (\ref{gen-n}).

The correspondence principle (25.$b$) is proved by straightforward
verification with the help of eq.(\ref{lift}). Finally, to prove
the normalization condition (25.$c$) it suffices  to note that
$\sigma(1)=1$. q.e.d.

\vspace{5mm} \noindent {\it Remark 1}. The constructed
$\ast$-product has a rather special structure: the bi-diffrential
operators at each power of $\hbar$ are determined by repeated
differentiations along the vector fields $X_i$,
\begin{equation}\label{Dn}
D_n(a,b)=\sum_{k,l\leq n}D_n^{i_1\cdots i_kj_1\cdots
j_l}(X_{i_1}\cdots X_{i_k}a)( X_{j_1}\cdots X_{j_l}b)\,,
\end{equation}
where the coefficients $D_n^{i_1\cdots i_kj_1\cdots j_l}\in {\Bbb
C}$ are universally expressed via the structure constants
$f_{ij}^k$ and $r_{ij}$, irrespective of the concrete
representation $\rho: X_i\rightarrow {\rm Vect}(M)$. For this
reason one may refer to this $\ast$-product as the universal
quantization formula. It would be very interesting to compare this
result with the Kontsevich quantization formula \cite{Kontsevich} and/or
give its ``sigma-model interpretation" \cite{CF1}.

\vspace{5mm}

\noindent
{\it Remark 2}. The above quantization admits a natural deformation of the
algebraic data entering to the $\ast$-product construction. Namely, the
cocycle $r$ may be replaced by any formal series
\begin{equation}  \label{def}
r^{\prime}_{ij}=r_{ij}+\hbar r^{(1)}_{ij}+\hbar^2r^{(2)}_{ij}+\cdots\,,
\end{equation}
where $r^{(k)}$ are arbitrary 2-cocycles of the algebra $L$. Note
that the matrix $r^{\prime}$ is formally invertible since $r$ is
so. As the result we get another star-product having the same
quasi-classical limit (25.$b$). The question arises as to whether
the deformation (\ref{def}) leads to essentially different
quantization or an equivalence transform can be found to
establish an isomorphism between two algebras of quantum
observables. In the next section we formulate the necessary and
sufficient conditions for two star-products, obtained in such a
way, to be equivalent.

\section{Two theorems of equivalence}

Given a QFL algebra $L$, consider two star-products $\ast$ and
$\ast^{\prime}$ associated to non-degenerate 2-cocycles $r$ and $
r^{\prime}$ of the form (\ref{def}). Following
\cite{Dr83}, we call these star-products to be {\it equivalent}
if there exists an invertible  operator
\begin{equation}
B:C^{\infty}(M)[[{\hbar}]]\mapsto C^{\infty}(M)[[{\hbar}]],  \label{eqop}
\end{equation}
establishing isomorphism of two star-algebras, i.e.
\begin{equation}
B(a\ast' b)= (Ba \ast Bb)  \label{inter}
\end{equation}
The most general expression for such an operator $B$ preserving
the peculiar structure of $\ast$-product (see Remark 1 of the
previous section) looks like
\begin{equation}  \label{b}
B=1+\hbar B_1+\hbar^2B_2+\,\cdots\,,
\end{equation}
where $B_k$ are finite order differential operators of the from
\begin{equation}  \label{bk}
B_k=\sum_{n=1}^{N}C^{i_1\cdots i_n}X_{i_1}\cdots
X_{i_n}\,\quad\quad C^{i_1\cdots i_n}\in {\Bbb C}\,.
\end{equation}

In the paper by Drinfeld \cite{Dr83} the theorem was stated that
the inequivalent Universal Deformation Formulas are classified by
the formal series in $\h$ with values in the second cohomologies
of the Lie algebra. However the proof of the theorem
was given neither in this paper \cite{Dr83} nor in any
of consequent papers. Here we present a simple proof
of the classification theorem for
the inequivalent Universal Deformation Formulas, obtained
by our procedure.
\vspace{0.5cm}

\noindent {\bf Theorem 3.} {\it Two star-products $\ast$ and
$\ast^{\prime}$ associated to formal 2-cocycles $r$ and
$r^{\prime}$ are equivalent if and only if}

\vspace{2mm}

$i)$ $r =r^{\prime}\,\, ({\rm mod}\,\, \hbar)$;

\vspace{2mm}

$ii)$ $r-r^{\prime}$ {\it is a coboundary }.

\vspace{0.5cm}

\noindent {\it Proof.} The necessity of the above conditions can
be proved by standard deformation quantization arguments based on
the consideration of Hochschild cohomologies. So we omit this
part of the proof and turn to the sufficiency.
By assumption of the theorem we have $r_{ij}-r^{\prime}_{ij}=f_{ij}^k\xi
_k$ , where $\xi _i={\hbar }\xi _i^{(1)}+{\hbar }^2\xi
_i^{(2)}+\,\,\ldots $ is a formal vector. Then the Lie algebra
isomorphism $Q: L_c\rightarrow L_c^{\prime }$, defined on the
basis elements as
\begin{equation}
Q(y_i)=y_i+\xi _ic\,,\quad \quad Q(c)=c\,
\end{equation}
induces isomorphism of the universal enveloping algebras $U(L_c)$ and $
U(L_c^{\prime })$, which then extends to isomorphism of associative algebras
$({\cal A}[[\hbar ]],\circ )$ and $({\cal A}[[\hbar ]],\circ ^{\prime })$.
This means that the operator
\begin{equation}
Qa(y,{\hbar })=a(y+\xi,\hbar )  \label{Q}
\end{equation}
intertwines the circle-products $\circ $ and $\circ ^{\prime }$ constructed
by $r$ and $r^{\prime }$ respectively. Since
\begin{equation}
D_i^{\prime }a=X_ia+\frac 1{i\hbar} [y_i\,,a\,]^{\prime
}=QDQ^{-1}\,, \quad \quad [a,b]^{\prime }=a\circ ^{\prime
}b-b\circ ^{\prime }a\,,  \label{DC}
\end{equation}
$Q$ maps $D$-constant elements to $D^{\prime}$-constant ones.
Then it is easy to see that operator
\begin{equation}
Ba=\pi Q\sigma (a)  \label{eqiop}
\end{equation}
has the form (\ref{b},\ref{bk}) and intertwines the star-products
$\ast$ and $*^{\prime }$. Indeed,
$$
(Ba)*^{\prime }(Bb)=(\pi Q\sigma (a))*^{\prime }(\pi Q\sigma
(b))=\pi (Q\sigma (a)\circ ^{\prime }Q\sigma (b))=
$$
$$
=\pi Q(\sigma (a)\circ \sigma (b))= B\pi (\sigma (a)\circ \sigma
(b))=B(a*b)
$$
q.e.d.

\vspace{5mm} So far the role of QFL algebra underlying the
quantization was twofold. First, it had appeared as the Lie
algebra of vector fields entering to the $r$-bracket construction
(\ref{RB}). Second, it had defined the fiber-wise $\circ$-product
on the auxiliary bundle of the formal universal enveloping algebra
$U(L_c)[[\hbar]]$. Now we are going to separate these functions to
obtain a more flexible formalism which, as we hope, admits a
generalization beyond the case of Lie algebras (see Sec. 6). This
may also be viewed as a further deformation of the fiber-wise
$\circ$-product giving an equivalent $\ast$-product on the base
manifold $M$.

 \vspace{0.5cm}
\noindent {\bf Theorem 4}. {\it Let }$(L,r)${\it \ and
}$(L^{\prime },r^{\prime })$ {\it be two QFL algebras
with non-degenerate 2-cocycles. If }$\dim L=\dim L^{\prime }${\it
\ then the associative algebras} $U(L_c)[[\hbar ]]${\it \ and
}$U(L_c^{\prime })[[\hbar ]]$ {\it \ are isomorphic. }

\vspace{0.5cm}

\noindent {\it Remark}. Without loss of generality we may assume that
$r=r'$ (if not, choose another basis in the Lie algebra). Then it
is sufficient to construct such an isomorphism $\phi:
U(L_c)[[\hbar ]] \rightarrow U(L_c^{\prime })[[\hbar ]]$ for the
case when algebra $L$ is fixed. For definiteness, we take $L_c$
to be the Heisenberg-Weyl algebra:
\begin{equation}  \label{GW}
\lbrack z_i,z_j]=i\hbar r_{ij}c\,,
\end{equation}
whose  formal universal enveloping algebra, denoted by $W$,
consists of formal power series in $z_i$ and $\hbar$ with complex
coefficients
\begin{equation}\label{GW1}
  a(z,\hbar)=\sum_{k,n\geq 0}\hbar^{k}a_k^{i_1\cdots i_n}z_{i_1}\cdots
  z_{i_n}\,.
\end{equation}
Since the underlying QFL algebra $L$ is commutative the matrix
${\cal R}$ entering to the definition (\ref{circ}) of
$\circ$-product is equal to unit and we arrive at the usual
Weyl-Moyal formula
\begin{equation}
a\circ b
=a\left(z_i+\left(\frac{i\hbar}{2}\right)r_{ij}\frac{\partial}{\partial
z_j }\right)b(z)={\rm
exp}\left(\frac{i\hbar}{2}r_{ij}\frac{\partial}{\partial z_i}
\frac{
\partial}{\partial w_j}\right)a(z)b(w)|_{z=w}\,.
\end{equation}
In what follows we will refer to $W$ as the Weyl algebra.

Now let $L$ and  $L^{\prime}$ be arbitrary QFL algebras of equal
dimension, then the stated isomorphism $\phi:
U(L_c)[[\hbar]]\rightarrow U(L'_c)[[\hbar]]$ can be  written as
the composition $\phi=\rho\rho'^{-1}$ of isomorphisms $\rho:
W\rightarrow U(L_c)[[\hbar]]$ and $\rho': W\rightarrow
U(L^{\prime}_c)[[\hbar]]$.

\vspace{0.5cm} \noindent {\it Proof}. Prescribing the degrees to
the formal variables: $\deg \ y_i=1$ and $\deg \ \hbar =2$, we
turn $W$ to a graded associative algebra assigned by the natural
filtration $W\supset W_1\supset W_2\ldots $ with respect to the
total degree $2k+n$ of the series terms (\ref{GW1}). The desired
isomorphism $\rho: W\rightarrow U(L_c)[[\hbar ]]$ is then
completely determined by its action on the generators $z_i$ of the
Heisenberg-Weyl algebra. Having in mind the invertibility of the
map $\rho$, we are looking for $\rho$ of the form
\begin{equation}  \label{anzats}
y_i=\rho (z_i)=z_i+Y_i(z,\hbar)\,,
\end{equation}
where $Y_i\in W_2$. Substituting this anzatz  to the commutation
relation of the algebra $L_c$ we get
$$[\rho (z_i),\rho  (z_j)]=i\hbar (f_{ij}^k\rho (z_k)+r_{ij})\,,
$$
or explicitly
\begin{equation}  \label{MC}
\partial _iY_j-\partial _jY_i=f_{ij}^k(z_k+Y_k)-\frac 1{i\hbar}
[Y_i,Y_j]\,,
\end{equation}
where
$$\partial _i =r_{ij}\frac{\partial}{\partial z_j}\,. $$
Let us show that the last equation has a unique solution for
$Y_i\in W_2$  subject to additional condition
\begin{equation}  \label{con}
z_ir^{ij}Y_j(z,\hbar )=0\,
\end{equation}
This in particular implies $Y_i(0,\hbar )=0$. Consider an
expansion of $Y_i$ in the homogeneous components
$$Y_i=\sum\limits_{k\geq 2}Y_i^{(k)}\,, \quad\quad \deg \
Y_i^{(k)}=k\,.$$ Substituting this expansion back to the equation
(\ref{MC}) we obtain a chain of equations
\begin{equation}\label{Y}
\begin{array}{l}
\partial_iY_j^{(2)}-{\partial}_jY_i^{(2)}=f_{ij}^kz_k,
\\[3mm]\displaystyle
\partial _iY_j^{(n+1)}-\partial _jY_i^{(n+1)}=f_{ij}^kY_k^{(n)}-\frac
1{i\hbar} \sum\limits_{k=2}^n[Y_i^{(n+2-k)},Y_j^{(k)}]\ ,\qquad
n>2\,.
\end{array}
\end{equation}
Treating now $Y_i^{(n)}$ as 1-forms on a linear space with
coordinates $z^i=r^{ij}z_j$ we see that above equations have a
structure $dY^{(n)}=F^{(n)}$, where 2-forms $F^{(n)}$ are known. A
necessary and sufficient condition for these equations to be
solvable is $dF^{(n)}=0$ (the Poincare' lemma). The last condition
can be proved by induction over degree $n$. For $n=2$ the r.h.s.
of the first line in (\ref{Y}) is obviously closed in view of the
cocycle condition $ f_{ij}^lr_{lk}+cycle(i,j,k)=0$. Suppose now
that $dF^{(n)}=0$ and $Y^{(n)}$ is a solution to equation
$dY^{(n)}=F^{(n)}$. Then it is not hard to check that
\begin{equation}
(dF^{(n+1)})_{ijk}=f_{ij}^l(dY^{(n)}-F^{(n)})_{kl}+cycle(i,j,k)=0\,.
\end{equation}
Thus, using the induction, we obtain that $dF^{(n)}=0$, at any
$n$. By the Poincare' lemma there exists a unique 1-form
$Y^{(n)}=d^{-1}F^{(n)}$ vanishing on the linear vector field
$v=z^i\partial /\partial z^i$ (condition (\ref{con})). The
explicit expression reads
\begin{equation}
Y_i^{(n)}(z)=\int\limits_0^1dtF_{ij}^{(n)}(tz)z^j\,.
\end{equation}

The inverse map $\rho^{-1}: U(L_c)[[\hbar]]\rightarrow W$ is
obtained by iterating equation
\begin{equation}
z_i=y_i-Y_i(z,\hbar)\,
\end{equation}
with respect to $z_i$. Since ${\rm deg}\,Y_i\geq 2$ these
iterations converge to the unique solution $z_i=\rho^{-1}(y_i)\in
W$. q.e.d.

Using this theorem we can reformulate the quantization procedure
of Sec. 2 in terms of Weyl algebra bundle for any QFL algebra $L$
of vector fields $X_i\in {\rm Vect}(M)$. Namely, starting with
the associative algebra $A[[\hbar]]=C^{\infty}(M)\otimes W$ we
introduce a set of its derivatives $D_i:A[[\hbar]]\rightarrow
A[[\hbar]]$,
\begin{equation}
  D_ia= X_ia+\frac1{i\hbar}[y_i,\,a]\,,
\end{equation}
In view of the Theorem 4, one can  choose the element
$y_i=z_i+Y_i(z)$ to satisfy the commutation relations for the Lie
algebra $L_c$. Upon this choice for $y$'s the derivatives $D_i$
form the Lie algebra $L$ with respect to the commutator. Now,
repeating the proof of the Theorem 1 one can show the existence
of an isomorphism between the subspace of $D$-constant elements in
${\cal A}[[\hbar]]$ and the space of quantum observables
$C^{\infty}(M)[[\hbar]]$. The pull-back of $\circ$-product on
${\cal A}[[\hbar]]$ via this isomorphism induces $\ast$-product
on $C^{\infty}(M)[[\hbar]]$. The latter is obviously equivalent
to the $\ast$-product from Sec. 2.

\section{Nonabelian example}

The simplest example of the Frobenius Lie algebra is provided by
two-dimensional Borel algebra $B$ with the Lie bracket
\begin{equation}\label{borel}
  [H,E]=E\,.
\end{equation}
The coadjoint action of the algebra $B$ on its dual space
$B^{\ast}$ is generated by the pair of linear vector
fields\footnote{As is seen, the carrier space $B^*$ stratifies on
two coadjoint orbits of dimension 0 and 2.  The former coincides
with the origin $0\in B^*$, while the latter is given by $B^*
-\{0\}\sim S^1\times {\Bbb R }^1$. The existence of a coadjoint
orbit of dimension equal to the dimension of the corresponding
Lie algebra is a general characteristic property of the Frobenius
algebras \cite{El}.}
\begin{equation}\label{EandH}
H=y\partial_y\,,\quad\quad E=y\partial_x\,,
\end{equation}
where $(x,y)$ are coordinates on $B^*\sim {\Bbb R}^2$. This yields
a quadratic Poisson brackets on ${\Bbb R}^2$
\begin{equation}\label{ehv}
\omega =E\wedge H =y^2\partial_x\wedge\partial_y\,.
\end{equation}
In principle, this  bracket can be quantized by means of the
general iterative procedure described in the previous sections.
In this simple case, however, there is a more direct way to obtain
the respective $\ast$-product. Namely, observe that the above
Poisson bivector can also be written as the wedge product of the
following vector fields:
\begin{equation}
X=\partial_x\,, \quad \quad Y=y^2\partial_y\,,
\end{equation}
so that
$$
\omega=X\wedge Y\,,\quad \quad [X,Y]=0\,.
$$
As the vector fields commute, the bracket can easily be quantized
by the Weyl-Moyal like formula
\begin{equation}\label{com}
  f\ast
  g=\sum_{n=0}^{\infty}\left(-\frac{i\hbar}{2}\right)^n\frac{1}{n!}
  (X^nf)(Y^ng)
\end{equation}
On the other hand, it is easy to prove by induction that
\begin{equation}
(y^2\partial)^n=y^n\prod_{k=0}^{n-1}(y\partial_y+k)
\end{equation}
Using this identity we can rewrite the above $\ast$-product in
terms of noncommuting vector fields (\ref{EandH}) as
\begin{equation}\label{EH}
f\ast g=\sum_{n=0}^{\infty}\left(-
\frac{i\hbar}{2}\right)^n\frac{1}{n!} (E^nf)(H^{(n)}g)\,,
\end{equation}
where we put
$$
H^{(n)}=H(H+1)\cdots(H+n-1)\,.
$$
Actually, expression (\ref{EH}) is the universal quantization
formula as it gives an associative $\ast$-product for any pair of
vector fields $E$ and $H$ realizing the Borel algebra
(\ref{borel}). This may be seen from the following line of
reasons. The vector fields (\ref{EandH}) induce a special
representation $\rho: U(B)\rightarrow D({\Bbb R}^2)$ of the
universal enveloping algebra $U(B)$ in the algebra of
differential operators on plane $D({\Bbb R}^2)$. Since $H$ and
$E$ are linearly independent with functional coefficients this
representation is exact, i.e. $\ker \rho = 0$
\footnote{Obviously, this property is valid for the coadjoint
action of any Frobenius Lie algebra $L$ since there is an open
set of points (the orbit of maximal dimension) at which the
corresponding vector fields are linearly independent.}. In other
words, there are no algebraic relations among the first-order
differential operators $E$ and $H$ other than (\ref{borel}) and
its algebraic consequences. On the other hand, the associativity
condition for the $\ast$-product (\ref{EH}), being written
explicitly, has a form of algebraic equations for the generators
$E$ and $H$ and, as we have just argued, they should be satisfied
as consequence of the commutation relations of the Borel algebra
only. Thus the associativity holds for any representation $\rho :
B\rightarrow {\rm Vect}({\Bbb R }^2)$.

 Originally, the Universal
Deformation Formula (\ref{EH}) was obtained in the paper
\cite{CGG}. The skew-symmetric form of the above quasi-exponential
formula (that corresponds to the Weyl ordering) was derived in
\cite{UDF}.

The formula (\ref{EH}) allows one to quantize an interesting
class of irregular Poisson brackets on ${\Bbb R }^2$. Any bracket
on plane is given by a single function
\begin{equation}\label{f}
\omega =f(x,y)\partial _x\wedge \partial _y
\end{equation}
The function $f$ is said to be {\it quasi-homogeneous of weight}
$\la$ upon the weights of $x$ and $y$ equal to $\alpha$ and $\beta $
respectively if
 it is an eigen-vector to the Euler operator $v$ with the
eigen-value $\la$,
\begin{equation}
vf=\la f,\qquad v=\alpha x\partial _x+\beta y\partial _y\,.
\end{equation}
In what follows we assume $\la\neq 0$. Introduce a Hamiltonian
vector field $u$ with respect to the canonical Poisson bracket on
${\Bbb R}^2$,
\begin{equation}
u=\partial _xf\partial _y-\partial _yf\partial _x
\end{equation}
Then it is easy to check that
\begin{equation}
\omega =\frac{v\wedge u}{\la},\qquad [v,u]=(\la-\alpha -\beta )u
\end{equation}
If $\alpha +\beta =\la$ the bivector (\ref{f}) is representable as
exterior product of the commuting vector fields $u$, $v$,
otherwise the vector fields form (after an obvious redefinition)
the Borel algebra (\ref{borel}). Thus, every quasi-homogeneous
Poisson bracket on plane can be quantized in a pure algebraic
manner by formulas (\ref{com}) or (\ref{EH}).

For example, let $(m,n,k,l)$ be a set of four integers, then
function
\begin{equation}
f=x^my^n + x^ky^l\,
\end{equation}
is quasi-homogeneous of weight $\la= nk-lm$ with $\alpha =n-l$ and
$\beta =k-m$. Conversely, given  the integer weights $\la, \alpha,
\beta$, then the most general quasi-homogeneous, analytical at zero,
function has the form
\begin{equation}
f(x,y)=\sum_{k} a_kx^{n_k}y^{m_k}\,,
\end{equation}
$a_k$ being  arbitrary constants and the sum goes over all
non-negative integers solutions $(n_k, m_k)$ to the linear
Diophantus equation
\begin{equation}
\alpha n +\beta m =\la\,.
\end{equation}
Depending on $\alpha$, $\beta$ and $\la$ this equation may have
finite or infinite number of non-negative solutions, or may have
none.

\section{Quantization of triangular Lie bialgebras}

In this section we briefly discuss how the output algebraic
structures entering the quantization formula for $r$-bracket
(\ref{Dn}) can be interpreted in the formal language of the Lie
bialgebra quantization. Here, we do not want to go into detailed
definition of all the related mathematical constructions but only
recall some basic points, useful for our purposes. More details
can be found, for example, in \cite{Takh}, \cite{UDF}.

As it was first shown by Drinfeld \cite{Dr83} the deformation
problem for the triangular Lie bialgebra $(L,r)$ is equivalent to
constructing the Universal Deformation Formula. The latter can be
immediately read off from the explicit expression for the
$\ast$-product (\ref{star}, \ref{Dn}) if one replace the vector
fields $X_i$ by the Lie algebra generators $y_i$. This leads to
the following \textit{universal twisting element}:
\begin{equation}
F=I\otimes I+\sum_{n,m,k=1}^{\infty} \h^n D_n^{i_1\ldots i_m
j_1\ldots j_k} y_{i_1}\ldots y_{i_m} \otimes y _{j_1}\ldots
y_{j_k}\in U(L)\otimes U(L)[[\hbar]]\,,
 \label{ON}
\end{equation}
$I$ being the unit in $U(L)$. The element $F$ possesses special
algebraic properties (to be specified further) resulted  from the
associativity and normalization conditions for the respective
$\ast$-product, which, in turn, are based essentially on the
Leibnitz rule for the derivatives $X_i(ab)=X_i(a)b+aX_i(b)$ and
the ``null on constant condition" $X_i(1)=0$. The algebraic
formalization for the last two relations is naturally achieved
by introducing the co-algebra structure on the universal
enveloping algebra $U(L)$, i.e. the co-multiplication $\triangle
: U(L)\rightarrow U(L)\otimes U(L)$ and co-unit $\varepsilon :
U(L)\rightarrow {\Bbb C}$ mappings. Being homomorphisms of the
associative algebra $U(L)$ with unit $I$, these operations are
completely determined by their action on the generators $y_i$ and
$I$:
\begin{equation}\label{coumnoz}
\begin{array}{cc}
\triangle y_i=y_i\otimes I+I\otimes y_i  \; ,& \qquad  \triangle I
= I \otimes I  \; ,\\ [3mm]
 \varepsilon ( y_i) =0  \; , &\qquad
\varepsilon ( I ) =1 \; .
\end{array}
\end{equation}
Now the associativity $(a* b) *c = a * (b * c)$ and normalization
condition (25.$c$) are encoded in the following relations for the
universal twisting element $F$:
\begin{equation}\label{cocyc}
\begin{array}{c}
\left( (\triangle \otimes I) \, F \right) \, (F\otimes I) =
\left( (I\otimes \triangle) \, F \right) \, (I \otimes F) \; , \\
[3mm]
 (\varepsilon \otimes I) F=(I\otimes \varepsilon) F=I \; ,
\end{array}
\end{equation}
while the correspondence principle (25.$b$) looks like
\begin{equation}\label{qclass}
F=I\otimes I-\frac{i\h}2\, r^{ij}e_i\otimes e_j\,\,\,\, {\rm mod}
\,\,(\h^2) \; .
\end{equation}
Another natural operation $S: U(L)\rightarrow U(L)$, called
antipode, is induced by the involutive anti-homomorphism of the
Lie algebra $L$:
\begin{equation}\label{counit}
  S(y_i)=-y_i\,, \quad S(I)=I\,,\quad S^2={\rm id}\,.
\end{equation}

The collection of the operations  $(\triangle, \varepsilon, S)$
endows $U(L)$ with a structure of the Hopf algebra
defined by three axioms:

\vspace{3mm}
 $i)\,\, (\triangle\otimes I)\triangle
=(I\otimes \triangle)\triangle$ - {\it co-associativity};

\vspace{3mm}
$
 ii)\,\,(\varepsilon \otimes I)\triangle =I=(I\otimes
\varepsilon)\triangle;
$

\vspace{3mm}
 $iii)\,\, m(S\otimes I)\triangle = u\circ \ve =m(I\otimes S)\triangle\,\,\,$
 ($\circ$ {\it stands for the composition of maps}).

\vspace{3mm}
 \noindent
 Here $m$ denotes the standard
multiplication in $U(L)$ and the map $u$ sends $c\in {\Bbb C}$ to
$c\cdot I\in U(L)$. If in addition $\triangle =\tau \triangle$,
where automorphism $\tau : a\otimes b \rightarrow b\otimes a$
permutes the multipliers in the tensor square $U(L)\otimes U(L)$,
then the Hopf algebra is called co-commutative\footnote{Omitting
the antipode operation $S$ and the related axiom (iii) one
arrives at the notion of an associative bialgebra. }.

Thus we arrive at the one-to-one correspondence between the
variety of $\ast$-products of the form (\ref{star}) and solutions
to the equations (\ref{cocyc}) subject to the ``boundary
condition" (\ref{qclass}); in doing so, the invariance of
equations (\ref{cocyc}) under the transformation
$$
F \longrightarrow F'=(\triangle \, B) \, F \, (B^{-1} \otimes
B^{-1}) \, ,\quad B \in U(L)[[\hbar]]\,,  \quad \varepsilon B =
I\,,
$$
corresponds the to equivalence of two such $\ast$-products
\cite{Dr83}. As another example of invariance we present the following discrete
symmetry:
\begin{equation}
\label{ON1}
F \rightarrow \widetilde{F} = \left[ (S \otimes S) \,
\tau F\tau \right]^{-1}
\end{equation}
Using the new solution $\widetilde{F}$ to the equations
(\ref{cocyc}), having the same quasi-classical limit
(\ref{qclass}), one can construct another $\ast$-product which is
not in general equivalent to the initial one (\ref{ON}) in the
sense of (\ref{inter}). It would be interesting to explicitly
compute the characteristic class of the respective $\ast$-product.

In order to illustrate how the universal deformation formula
gives rise  to the deformation of the Lie bialgebra structure,
we remind the construction of the quantum universal enveloping
algebra $U_{\hbar}(L)$. As an associative algebra, $U_{\hbar}(L)$
coincides with $U(L)[[\hbar]]$. The respective co-unit
$\varepsilon$ remains the same as in (\ref{coumnoz}), while the
co-multiplication and the antipode for $U_{\hbar}(L)$ are
obtained from those of $U(L)$ by the twisting transformation
\begin{equation}
\begin{array}{cc}
\triangle_{\hbar} a= F^{-1}(\triangle a) F\,,& \qquad S_{\h}
a=U^{-1} S a  U\,,\\ [3mm] U=m((S\otimes I) F)\,, &\qquad a\in
U_{\hbar}(L)\,.
\end{array}
\end{equation}
The validity of the Hopf algebra axioms for the deformed
co-algebra structure can be verified with the help of conditions
(\ref{cocyc}).

Given the associative bialgebra $U_{\hbar}(L)$, one can twist the
multiplication of any $U_{\hbar}(L)$-module algebra or
co-algebra, in particular, to write an explicit symbol
realization for the deformed coordinate rings of various algebraic
varieties \cite{UDF}. Among the most important examples of such a
kind it is worth to mention the Takhtajan approach to the quantum
group construction \cite{Takh}. Let $G$ be the Lie group
corresponding to the Lie algebra $L$, and denote by $G_{\hbar}$
corresponding quantum group. The latter is defined in terms of an
associative algebra of functions ${\rm Fun} (G_{\hbar})$. As a
linear space, the algebra ${\rm Fun}(G_{\hbar})$ is identified
with $C^{\infty}(G)[[\hbar]]$ and equipped with the following
star-product:
\begin{equation}
a\star b = M (F^L (F^{-1})^R (a\otimes b))\,, \label{starr}
\end{equation}
where $M$ stands for the ordinary commutative multiplication in
$C^{\infty}(G)$ and $F^{L}$, $(F^{-1})^{R}$ denote the twisting
elements $F$, $F^{-1}$ in the representation of the left $\{
X^L_i \}$ and anti-representation of the right $\{ X^R_i \}$
invariant vector fields on the Lie group $G$, respectively. The
associativity of $\star$-product follows from the first equation
in (\ref{cocyc}) and the fact that $X^R_i ~\mapsto~ X^L_i$ is
involutive anti-homomorphism of the Lie algebras of vector fields.
Evaluation of the quasi-classical limit for $\star$-product
(\ref{starr}) leads to the following Poisson bracket:
$$
\{ a , \, b \} = r^{ij} \, (X_i^L a \,\,  X_j^L b -  X_i^R a \,\,
X_j^R b )\,.
$$
As is seen this bracket is a particular example of $r$-brackets
considered in the previous sections.

Note that one can use  the transformed twisting element
(\ref{ON1}) or even a combination of two different elements
(\ref{ON}) and (\ref{ON1}) in the definition (\ref{starr}) to
obtain a different inequivalent quantizations. However the
special ``adjoint" structure of the deformation (\ref{starr})
(the ``right" twisting element is taken to be inverse to the
``left" one) implies also that the standard group
co-multiplication $\triangle : {\rm Fun}(G_{\hbar})\rightarrow
{\rm Fun}(G_{\hbar})\otimes { \rm Fun}(G_{\hbar})$ defined by the
relation
\begin{equation}
  (\triangle _G f)(g_1,g_2)=f(g_1g_2)\,
\end{equation}
is indeed a homomorphism of the algebra
$(C^{\infty}(G))[[\hbar]], \star )$ to
$(C^{\infty}(G)[[\hbar]], \star )\otimes
(C^{\infty}(G)[[\hbar]],\star )$.

Our explicit formulas, being combined with the Takhtajan
construction \cite{Takh}, provide a transparent and explicit description
for the quantum algebra of functions ${\rm Fun}(G_{\hbar})$, that can
be viewed as an alternative to the standard FRT method \cite{FRT}.

\section{Discussion and Conclusion}
To summarize, in this paper we have proposed a simple
quantization procedure for the Poisson brackets associated to the
classical triangular $r$-matrix and giving, as a byproduct, the
explicit quantization formula for Lie bialgebras.  Abstracting
from the peculiar form of the $r$-brackets (\ref{RB}),
(\ref{lalg}) one can find a noticeable similarity  between our
quantization algorithm and Fedosov's construction for symplectic
manifolds. Indeed, in both the cases one begins with an auxiliary
quantum  bundle assigned by a fiber-wise $\circ$-product; in so
doing, the space of quantum observables is identified to the
center of this algebra. At the next step a new nontrivial
embedding is constructed to induce the $\ast$-product on the
Poisson manifold as a pull-back of the $\circ$-one.  This is
achieved by introducing an appropriate Lie algebra of external
derivatives and identifying the image of the embedding map with
their kernel subspace (in fact $\circ$-subalgebra). Moreover, the
Theorem 4 of Sec. 3 suggests that, similar to the Fedosov
approach, we can always start with the quantum bundle of Weyl
algebras and construct the embedding map in an explicit recurrent
manner with any accuracy in the deformation parameter. In our
special algebraic situation, however, the use of the universal
enveloping algebra bundle seems to be more adequate to the
problem, considerably simplifying the whole construction. Another
novel point of our approach, looking from the pure algebraic
perspective,  is the non-commutativity of the aforementioned
derivatives for any choice of fiber-wise $\circ$-product. This is
a distinction from the Fedosov deformation quantization  which is
essentially based  on the notion of the  {\it Abelian} connection
corresponding, in this language, to a certain pair-wise commuting
set of derivatives associated with any symplectic connection.

Let us briefly discuss a possibility to extend our star product construction to a
more general class of Poisson manifolds than considered in this paper.  It seems
useful to seek for the generalizations making use of a natural relationship between
the Fedosov deformation quantization and BRST theory \cite{GL}
\footnote{ On the other hand,
the BRST language could be useful in attempts to apply the present
$\ast$-product construction to the field theory problems, as it provides
a uniform description for both deformation quantization of the phase space
observable algebra and gauge symmetry or/and  the Hamiltonian constraint algebra,
see \cite{BGL}. }.
The key idea behind the
identification of both the quantization schemes was a realization of the symplectic
manifold $M$ as a second class constrained surface embedded into cotangent
bundle $T^*M$.  Let us explain how this approach, originally
found for the symplectic manifold, can be first extended to the case
of $r$-brackets (\ref{RB}), (\ref{lalg}).
Consider a
trivial vector bundle ${\cal{M}}=M\times V$ over a smooth manifold $M$, assigned by
the following Poisson brackets
\begin{equation}\label{11}
\{ x^\mu ,x^\nu \}=0\,,\quad\{p_i,p_j\}=f^k_{ij}p_k + r_{ij}\,,\quad
\{p_i,x^\mu\}=X^\mu_i(x)\,.
\end{equation}
Here $x^\mu$ are local coordinates on $M$ and $p_i$ are linear
coordinates on $V$. It is easy to check that these brackets satisfy
the  Jacobi identity iff (i) the vector fields $X_i$ generate a
Lie algebra  (\ref{lalg})  and (ii) matrix $r$ satisfies
cocycle condition  (\ref{cc}). If in addition $\det(r_{ij})\neq 0$,
then the second class constraints $T_i=p_i\approx 0$ define an
embedding of $M$ into $\cal M$ as a zero section. The
corresponding Dirac bracket on $\cal M$ induces Poisson bracket
on $M$,
\begin{equation}\label{Dirac}
\{x^\mu ,x^\nu\}_{Dirac}=r^{ij}X^\mu_i(x)X^\nu_j(x)\,.
\end{equation}
This relation identifies the Hamiltonian theory on $M$ to the
second class constrained theory on $\cal M$. Note that in general
both the brackets (\ref{11}) and (\ref{Dirac}) are irregular,
that does not, however, obstruct the mentioned identification. The
common method of BRST treatment of the second class constrained
systems implies the conversion of the second class constraints
$T_i$ to the first class ones by means of a proper extension of
the phase manifold $\cal M$ (see \cite{BT}, and references
therein). The minimal possibility to do this is to duplicate the
vector bundle $V$, that is to consider the direct product ${\cal
M}'={\cal M}\times V$, where $V$ is a linear symplectic space
assigned by canonical Poisson bracket
\begin{equation}\label{111}
\{z_i, z_j\}=r_{ij}\,,
\end{equation}
$z_i$ being linear coordinates on the second copy $V$. Together
Rels. (\ref{11}) and (\ref{111}) define the Poisson structure on ${\cal M}'$
(we mean that bracket of $z$ are vanishing to other variables).
Now it is possible to extend the constraints $T_i$ from $\cal M$ to
${\cal M}^\prime $ getting an
equivalent first class constrained theory $T'_i=T_i+z_i+ o(z^2)$, where the
higher orders in the formal variables $z_i$ are determined from
the requirement of involution of the effective first class constraints
$T'_i$
\begin{equation}
 \{T'_i, T'_j\}=f_{ij}^kT'_k \,.
\end{equation}
The Poisson brackets (\ref{11}), (\ref{111}) on ${\cal M}'$ can be
easily quantized if one restricts the space of observables to the
functions on ${\cal M}'$ at most linear in $p_i$. The connection
with constructions of previous sections is established by the
relation $D_ia=[T_i,a]$, where $a=a(x,z,\hbar)$. The construction
of nilpotent BRST charge and identification of physical
observable algebra with the special BRST cohomology imply a
further enlargement of the phase space ${\cal M}'$ by the ghost
variables. For the case of $M$ being a symplectic manifold, this
BRST quantization programme has been implemented in \cite{GL},
and it seems no special modification is required in the case of
the Poisson manifold (\ref{RB}), (\ref{lalg}).

In this form the above quantization procedure may be generalized
to the case when all the structure constants $f^k_{ij}$, $r_{ij}$
entering to the fundamental Poisson brackets (\ref{11}),
(\ref{111}) are allowed to depend on the point $x\in M$ \cite{LS}.
 As a particular case this includes the quantization of dynamical
$r$-matrix \cite{XU} with the vector fields $X_i$ generating a
Lie bialgebroid.

\vspace{0.5 cm}

{\bf Acknowledgments.} We are thankful to I.A.~Batalin,
M.A.~Grigoriev, O.V.~Ogievetsky, M.A.~Olshanetsky, I.Yu.~Tipunin
for useful discussions.
We also acknowledge O.V.~Ogievetsky for the constructive criticism concerning
the preliminary version of this article.
The work of SLL is partially supported by
RFBR grant 00-02-17-956 and the grant INTAS 00-262. The work of
AASh is supported by RFBR grant 02-02-06879 and Russian Ministry
of Education under the grant E-00-33-184. The work of VAD is
partially supported by the grant INTAS 00-561 and by the Grant
for Support of Scientific Schools 00-15-96557. The work of API is
partially supported by the RFBR grant 00-01-00299. API is also
grateful to Max-Plank-Institute fur Mathematik in Bonn for kind
hospitality and support. SLL is thankful to Lars Brink and
Robert Marnelius for their warm hospitality at the Chalmers
University and to STINT (Swedish Foundation for International
Cooperation in Research and Higher Education) for the financial
support.

\end{document}